\documentclass[runningheads]{llncs}

\usepackage[T1]{fontenc}
\usepackage{graphicx}
\usepackage{booktabs}
\usepackage{url}
\usepackage{amssymb}
\usepackage{marvosym}

\newcommand{\cmark}{\ensuremath{\checkmark}}
\newcommand{\pmark}{\ensuremath{\triangle}}

\begin{document}

\title{PolyDebate: A Game-Orchestrated Multimodal System for Debate Skills Practice and Evaluation}
\titlerunning{PolyDebate}

\author{Jianing Yin\inst{1} \and
Weng Pan Kuan\inst{2} \and
Xiaoyun Liu\inst{1} \and
Zhiyuan Wen\inst{1,3}\textsuperscript{(\Letter)} \and
Yuxuan Li\inst{1} \and
Milos Stojmenovic\inst{4} \and
Jiannong Cao\inst{1,3}
}
\authorrunning{J. Yin et al.}
\institute{Department of Computing, The Hong Kong Polytechnic University, Hong Kong SAR, China\\
\email{\{jianing-laetitia.yin, xiaoyun.liu, yuuxuann.li\}@connect.polyu.hk}\\
\email{\{zhiyuan.wen, jiannong.cao\}@polyu.edu.hk}
\and
Department of Data Science and Artificial Intelligence, The Hong Kong Polytechnic University, Hong Kong SAR, China\\
\email{wengpan.kuan@connect.polyu.hk}
\and
Institute for Higher Education Research and Development, The Hong Kong Polytechnic University, Hong Kong SAR, China
\and
Department of Computer Science and Electrical Engineering, Singidunum University, Belgrade, Serbia\\
\email{mstojmenovic@singidunum.ac.rs}
}

\maketitle

\begin{abstract}
Debate is a structured form of persuasive communication that trains argument construction, rebuttal, oral delivery, and audience awareness. These skills are valued in education, language learning, and professional communication. Recent AI debate systems and LLM-based judges have advanced argument generation and debate evaluation, but most remain text-centered and rarely support learners through a complete multimodal practice experience. We introduce PolyDebate, a game-orchestrated multimodal system for English debate practice and evaluation. PolyDebate guides learners through staged one-on-one (1v1) debates with an AI opponent, while skill cards, props, and coins make persuasive strategies explicit and turn practice into a game-like interaction. During each session, the system captures learner speech and visual delivery evidence, generates context-aware opponent responses, and produces rubric-informed stage-level and overall feedback. PolyDebate is available as both an immersive Unity 3D game version and a web platform version that share the same workflow and evaluation services. Four studies covering AI opponent quality, evaluation coverage, AI judge feedback, and user perception show that PolyDebate brings debate interaction, gamified scaffolding, multimodal assessment, and structured feedback together in a practical workflow for debate skills practice. The demonstration video is available at \url{https://youtu.be/mHwBG1_8Ebk}.

\keywords{Debate practice \and Multimodal learning system \and AI debater \and Automated evaluation \and Game-based learning}
\end{abstract}

\section{Introduction}

Debate is a useful form of persuasive communication that asks learners to construct arguments, use evidence, respond to opposing views, and deliver speeches to an audience. It is widely used in education and language learning because it trains critical thinking, spoken communication, and audience awareness~\cite{lee2023chatgptdebate,cassim2025gai,park2025critical}. However, high-quality debate practice is hard to provide at scale: learners need a responsive opponent, a clear debate procedure, and formative feedback on both argument quality and oral delivery.

Large language models (LLMs) open new opportunities for AI-supported debate practice: AI debaters can retrieve evidence, generate arguments, and sustain multi-turn interactions~\cite{slonim2021autonomous,zhang2024agent4debate,aryan2024debatepartners}, and LLM-based judges can support scoring, critical-thinking assessment, and guided debate activities~\cite{liang2024debatrix,lee2023chatgptdebate,park2025critical}. Yet two limitations remain. First, most systems focus on generating strong arguments or scoring transcripts rather than supporting a complete learner-facing practice process, lacking staged guidance, strategy scaffolding, and immediate feedback. Second, debate is inherently multimodal: persuasiveness depends not only on argument content but also on speech delivery, facial expression, gesture, and posture, so text-only practice and evaluation lose important evidence about delivery.

To fill these gaps, we introduce \textbf{PolyDebate}, a game-orchestrated multimodal system for English debate practice and evaluation, built from three modules. The \textbf{Game-Orchestrated Practice Module} runs a complete 1v1 debate round with side assignment, stage control, speaking turns, skill cards, props, and coins. The \textbf{Stage-Aware AI Debater Module} generates opponent responses conditioned on the motion, debate side, current stage, interaction history, and assigned skill card, so the AI acts as both an opponent and a pedagogical example. The \textbf{Rubric-Aligned Multimodal Evaluation Module} analyzes transcript, speech/audio, and visual evidence to produce stage-level and overall feedback. PolyDebate is implemented as both an immersive Unity 3D game version and a web platform version that share the same workflow and evaluation services.

In summary, this paper contributes (i) PolyDebate, an integrated 1v1 debate-practice system combining staged oral debate, an AI opponent, two interface versions, and learner-facing feedback; (ii) a game-orchestrated practice mechanism with skill-card guidance, props, and coins for in-game support and immediate feedback; and (iii) a rubric-aligned multimodal evaluation workflow linking transcript, speech/audio, and visual evidence to actionable learning feedback.

\section{Related Work}

\subsubsection{AI debaters and LLM-based debate systems.}
Early autonomous debating systems showed that AI can retrieve evidence, construct arguments, and deliver rebuttals in competitive settings~\cite{slonim2021autonomous}. Agent4Debate uses multi-agent collaboration for argument preparation, analysis, writing, and review~\cite{zhang2024agent4debate}; DebateBrawl combines LLMs with genetic algorithms and adversarial search for adaptive arguments~\cite{aryan2024debatepartners}; and R-Debater improves multi-turn generation through retrieval-augmented argumentative memory~\cite{li2025rdebater}. These systems target AI debating performance rather than staged practice guidance, skill-card guidance, or multimodal learner feedback; PolyDebate instead uses the AI opponent within a learner-facing 1v1 practice workflow.

\subsubsection{Automated debate evaluation.}
Debatrix proposes a multi-dimensional LLM judge that analyzes performance across dimensions and chronological stages~\cite{liang2024debatrix}. Debatable Intelligence shows that evaluating debate speeches requires complex judgments about argument strength, relevance, organization, style, and tone~\cite{sternlicht2025debatable}. InspireDebate introduces InspireScore, combining subjective dimensions (e.g., emotional appeal, argument clarity) with objective ones (e.g., factual authenticity, logical validity)~\cite{wang2025inspiredebate}. Multi-agent debate chatbots have also assessed learners' critical thinking from their argumentation~\cite{park2025critical}. These methods, however, evaluate debate mainly through text and do not jointly cover argument content, oral delivery, visual behavior, and learner-facing feedback.

\subsubsection{AI-supported debate learning.}
A ChatGPT-based debate game shows how prompt engineering can support topic selection, debate flow, difficulty control, and simple scoring~\cite{lee2023chatgptdebate}. In EFL training, generative pedagogical agents and teacher-guided scaffolds have been used for role-driven debate preparation~\cite{cassim2025gai}, and structured chatbots can guide learners through evidence, warrant, and rebuttal stages, mainly for critical-thinking assessment~\cite{park2025critical}. Such systems usually cover only part of the practice experience and rarely combine a complete staged round, speech interaction, an AI opponent, skill cards, game mechanics, and multimodal feedback in one playable environment. PolyDebate addresses this gap through a single 1v1 workflow.

\section{PolyDebate}

\subsection{Overview and Workflow}

\begin{figure}[t]
    \centering
    \includegraphics[width=1\textwidth]{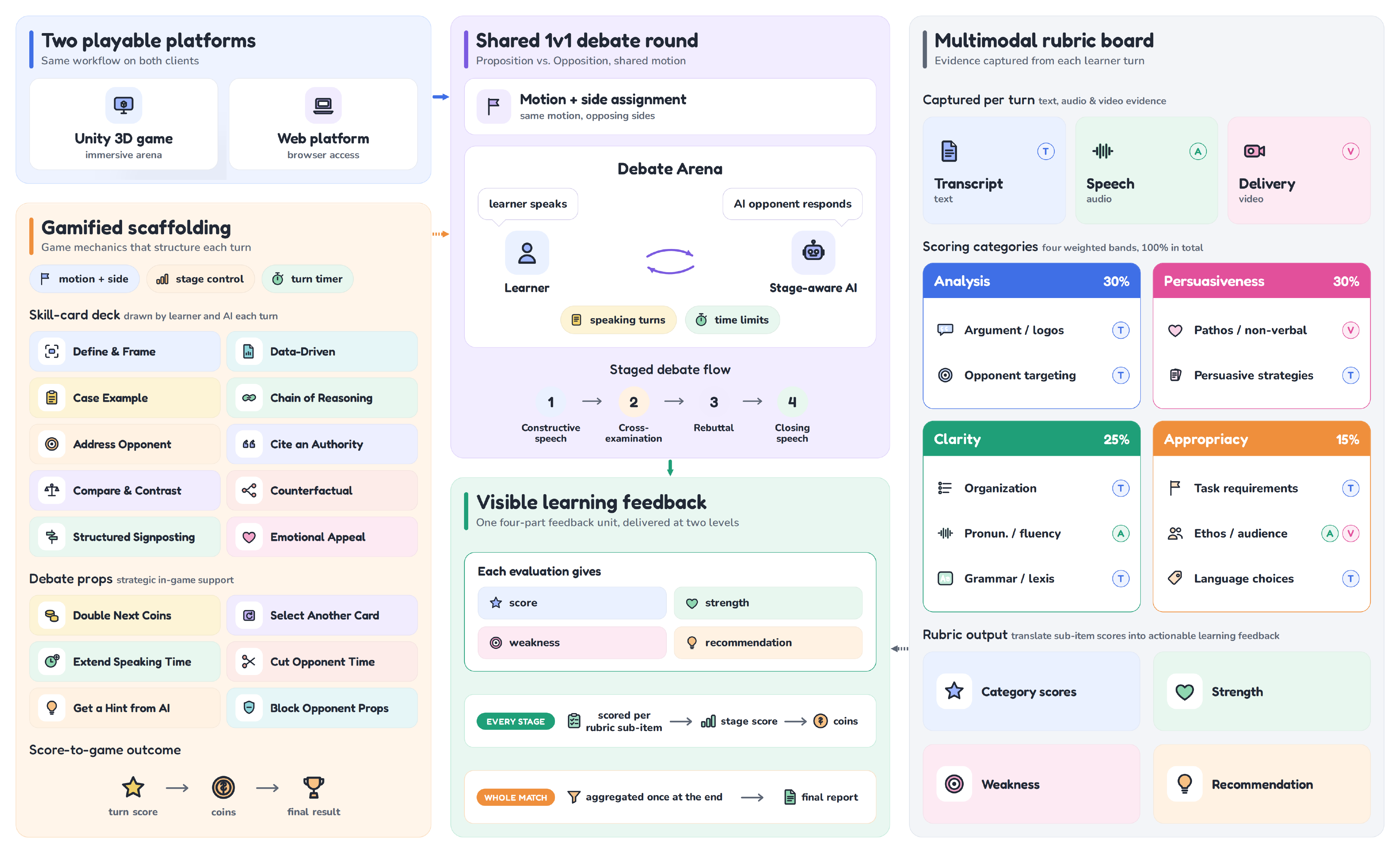}
    \caption{PolyDebate overview for gamified multimodal debate practice and rubric-aligned feedback.}
    \label{fig:workflow}
\end{figure}

PolyDebate combines staged oral debate, AI opponent interaction, rubric-aligned evaluation, and lightweight game mechanics in a single practice round, summarized in Fig.~\ref{fig:workflow}, and runs as a Unity 3D game version for immersive practice and a web platform version for browser access. Centered on a complete 1v1 round rather than an isolated chat, a session begins with a debate motion, assigns the learner and AI to opposing sides, and then follows a fixed four-stage sequence: constructive speech, cross-examination, rebuttal, and closing speech. Throughout, the system maintains a shared debate state (motion, sides, stage type, assigned skill cards, recent speeches, timing constraints, and game records) that the AI debater, the evaluation module, and the gamification layer all read from, keeping generation, assessment, and game feedback synchronized.

Each stage follows a repeated turn structure. The system assigns skill cards to the learner and the AI debater, shows stage-specific guidance, and lets the learner buy props before speaking. The learner's speech is captured and transcribed, with audio and video evidence prepared for analysis. The AI debater then produces an opposing turn appropriate to its side, stage, the learner's previous speech, and its assigned technique, and the evaluation module scores the turn under the four rubric categories, converting the score into coins and game feedback. After all stages, the turn records are aggregated into an overall evaluation with a final result, rating, strengths, weaknesses, and recommendations.

\subsection{Gamified Practice Design}

The gamification layer turns abstract debate strategies into explicit in-game guidance cues, choices, and outcomes. At the start of each stage, both the learner and the AI debater are assigned skill cards that specify concrete techniques (e.g., \emph{Data-Driven}, \emph{Chain of Reasoning}, \emph{Address Opponent}, \emph{Emotional Appeal}), so both sides of the exchange are shaped by explicit techniques rather than free-form conversation. Figure~\ref{fig:workflow} lists the full skill-card deck and props shop.

Coins and props provide immediate feedback and learner agency. After each evaluated turn the score is converted into coins, and cumulative coins decide the final outcome. Before the next stage, learners can spend coins in the props shop on limited support or strategic effects (the full set is listed in Fig.~\ref{fig:workflow}). These mechanics do not replace the rubric evaluation; they translate assessment results into visible progress and make repeated practice more engaging. Together, stage control, skill cards, AI opponent behavior, rubric-based scoring, and coin outcomes turn a debate round into a complete practice loop.

\subsection{AI Debater}

The AI debater is a stage-aware opponent rather than a general question-answering agent. For each AI turn, the generation module receives a structured context: the motion, both sides, the current stage, the learner's latest speech, the debate history, and the AI's skill card. This keeps the response side-consistent and stage-appropriate, for example, by establishing a position in the constructive stage, probing the opponent in cross-examination, attacking weak points in rebuttal, and summarizing the strongest clash points in closing. Because the same skill-card representation used by the learner is also applied to AI generation, the response is not only an opposing argument but also a visible example of a debate technique in context.

After text generation, the response is synthesized with Edge-TTS (a Microsoft neural English voice) and passed to the LiveTalking digital-human module, where a Wav2Lip-based lip-synchronization model~\cite{prajwal2020wav2lip} drives the avatar's mouth movements. The audio-visual stream is delivered over WebRTC, so the opponent appears as a speaking debate partner rather than a text-only agent, and each response is stored in the debate history for later turns and the final evaluation.

\subsection{AI Evaluation of Debate Performance}

The evaluation module is organized around a rubric adapted from ELC2012, an undergraduate English course on persuasive communication at The Hong Kong Polytechnic University. This course rubric provides the pedagogical basis for four categories: Analysis, Persuasiveness, Clarity, and Appropriacy. It anchors the AI evaluation in an existing teaching context and makes feedback more interpretable.

Following this rubric, PolyDebate maps multimodal evidence onto the same four categories, as detailed in the rubric board of Fig.~\ref{fig:workflow}. \textit{Analysis} draws on the transcript for argument/logos and opponent targeting; \textit{Persuasiveness} adds video evidence for non-verbal delivery; \textit{Clarity} adds audio evidence for pronunciation and fluency; and \textit{Appropriacy} adds audio and video evidence for audience appeal and delivery appropriateness. The categories are weighted 30\%, 30\%, 25\%, and 15\%, respectively.

The turn-level evaluator takes the stage, motion, learner side, transcript, opponent context, skill card, and audio/video descriptors, and returns category scores and item-specific feedback, which is converted into coins. A final evaluator then aggregates the debate history, multimodal evidence, and turn-level results into an overall rating, strengths, weaknesses, and recommendations, so feedback reflects the full trajectory rather than a single utterance.

\section{Demonstrations}

\begin{figure}[t]
    \centering
    \includegraphics[width=0.87\textwidth]{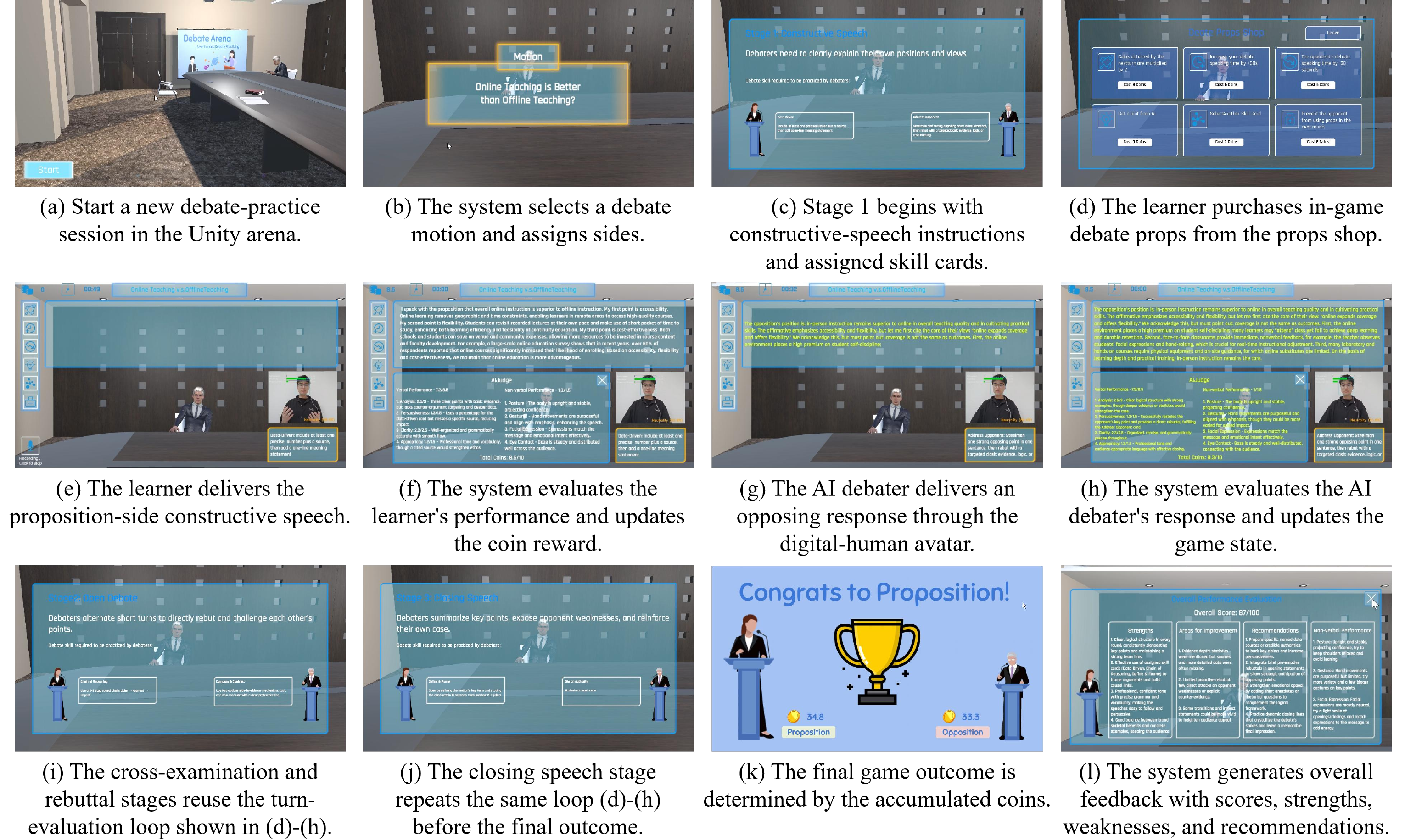}
    \caption{Demonstration walkthrough of a complete 1v1 PolyDebate session in the Unity 3D game version.}
    \label{fig:demonstration}
\end{figure}

\begin{figure}[t]
    \centering
    \includegraphics[width=0.87\textwidth]{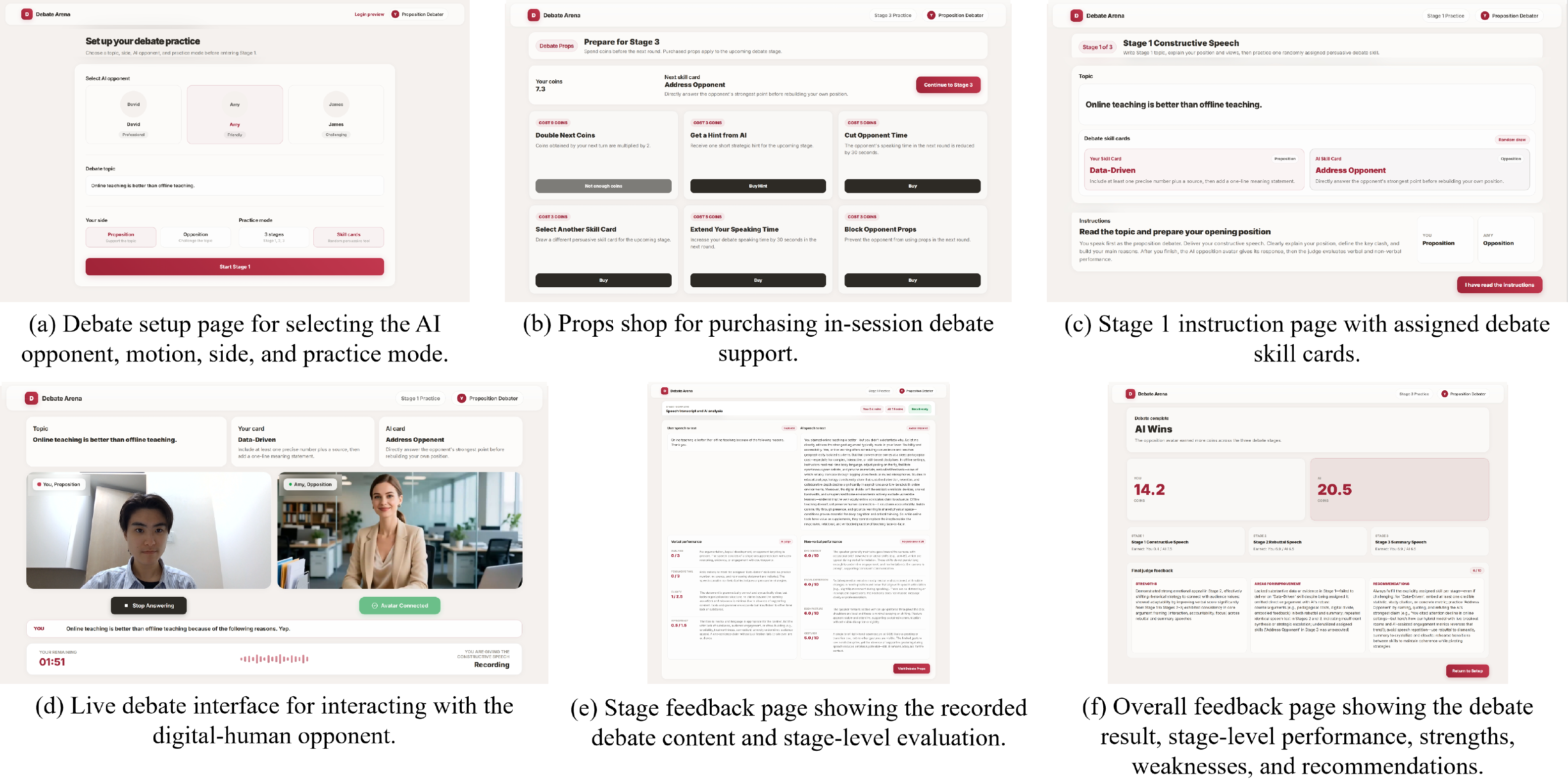}
    \caption{Representative key interfaces of the PolyDebate web platform version.}
    \label{fig:web-interfaces}
\end{figure}

Figure~\ref{fig:demonstration} walks through a complete 1v1 session in the immersive Unity 3D arena, with each panel showing one step of the practice-evaluation loop: from session start and side assignment, through the constructive, cross-examination, rebuttal, and closing stages, to the final coin outcome and overall feedback. Figure~\ref{fig:web-interfaces} shows representative screens from the browser-based web platform version, which shares the same workflow, and the accompanying video presents both versions at \url{https://youtu.be/mHwBG1_8Ebk}.

\section{Experiments and Evaluation}

We evaluated PolyDebate through four analyses: AI opponent next-turn response quality, evaluation coverage against representative frameworks, AI judge feedback quality with ablations, and a user-perception study.

\subsection{AI Opponent Next-Turn Response Quality}

This experiment evaluates the AI debater as a practice opponent: given a learner's previous turn and the current context, each system produces one opposing response. We sampled 100 next-turn cases from the Intelligence Squared Debates Corpus~\cite{zhang2016oxford} distributed with ConvoKit~\cite{chang2020convokit}. The cases cover eight public debates and were mapped to the four PolyDebate stages. Each case contains a human debate utterance treated as the learner-side turn, the opposite AI side, up to two previous turns as context, and an assigned skill card. All variants used the same GPT-5.4 mini model for generation. An independent LLM then scored the three anonymized responses per case on a 1--5 scale: six criteria adapted from InspireScore~\cite{wang2025inspiredebate} cover subjective and objective quality, and three skill-usage criteria check whether the target technique is visibly demonstrated as a pedagogical example. The overall score averages the subjective, objective, and skill scores.

Table~\ref{tab:ai-opponent-quality} reports the results. PolyDebate achieves the highest overall score (4.0) against the generic LLM opponent (3.1) and the stage-only opponent (3.6). The largest gain is in skill usage (2.1 to 3.9): PolyDebate visibly demonstrates target techniques within the opponent's response, showing that skill cards are more than interface guidance cues and also shape the AI into a stage-aware pedagogical example.

\begin{table}
\caption{AI opponent next-turn response quality. Scores use a 1--5 scale.}
\label{tab:ai-opponent-quality}
\centering
\scriptsize
\resizebox{\textwidth}{!}{%
\begin{tabular}{lccccccccccccc}
\toprule
 & \multicolumn{5}{c}{Subjective quality} & \multicolumn{3}{c}{Objective quality} & \multicolumn{4}{c}{Skill usage} &  \\
\cmidrule(lr){2-6}\cmidrule(lr){7-9}\cmidrule(lr){10-13}
\textbf{Variant} & \textbf{EA} & \textbf{AC} & \textbf{AA} & \textbf{TR} & \textbf{Avg.} & \textbf{FA} & \textbf{LV} & \textbf{Avg.} & \textbf{Pres.} & \textbf{Corr.} & \textbf{Vis.} & \textbf{Avg.} & \textbf{Overall} \\
\midrule
Generic LLM opponent & 3.2 & \textbf{4.0} & 3.2 & 3.8 & 3.6 & 3.6 & 3.9 & 3.8 & 2.0 & 1.8 & 2.6 & 2.1 & 3.1 \\
Stage-only opponent & 3.7 & \textbf{4.0} & 3.7 & 4.5 & 4.0 & 3.7 & 3.9 & 3.8 & 2.8 & 2.6 & 3.2 & 2.9 & 3.6 \\
PolyDebate opponent & \textbf{3.8} & \textbf{4.0} & \textbf{4.3} & \textbf{4.6} & \textbf{4.2} & \textbf{3.8} & \textbf{4.0} & \textbf{3.9} & \textbf{4.2} & \textbf{3.5} & \textbf{4.1} & \textbf{3.9} & \textbf{4.0} \\
\bottomrule
\end{tabular}}
\\[2pt]
\scriptsize EA=emotional appeal; AC=argument clarity; AA=audience adaptation; TR=topic relevance; FA=factual authenticity; LV=logical validity; Pres.=skill presence; Corr.=skill correctness; Vis.=pedagogical visibility.
\end{table}

\subsection{Debate Evaluation Coverage}

We compare PolyDebate's evaluation coverage with representative frameworks: Debatrix~\cite{liang2024debatrix}, InspireScore~\cite{wang2025inspiredebate}, Debatable Intelligence~\cite{sternlicht2025debatable}, and a multi-agent debate chatbot for critical-thinking assessment~\cite{park2025critical}. As shown in Table~\ref{tab:evaluation-coverage}, these methods evaluate debate mainly from text and cover aspects such as argument quality, relevance, organization, language use, and persuasive strength, but none jointly covers argument content, oral delivery, visual behavior, and learner-facing feedback together. PolyDebate adapts the ELC2012 rubric into a multimodal framework spanning text, audio, and video across analysis, persuasiveness, clarity, and appropriacy, and converts the results into structured feedback with strengths, weaknesses, and recommendations. This provides broader coverage and is better suited to formative learning.

\begin{table}
\caption{Comparison of representative debate and debate-context assessment frameworks in terms of modality, assessed aspects, and learner-facing feedback output.}
\label{tab:evaluation-coverage}
\centering
\scriptsize
\resizebox{\textwidth}{!}{%
\begin{tabular}{lcccccccccccccccc}
\toprule
 & \multicolumn{3}{c}{Modality} & \multicolumn{2}{c}{Analysis} & \multicolumn{2}{c}{Persuasiveness} & \multicolumn{3}{c}{Clarity} & \multicolumn{3}{c}{Appropriacy} & \multicolumn{3}{c}{Feedback output} \\
\cmidrule(lr){2-4}\cmidrule(lr){5-6}\cmidrule(lr){7-8}\cmidrule(lr){9-11}\cmidrule(lr){12-14}\cmidrule(lr){15-17}
\textbf{Framework} & \textbf{T} & \textbf{A} & \textbf{V} & \textbf{Arg.} & \textbf{Opp.} & \textbf{Pathos} & \textbf{Strat.} & \textbf{Org.} & \textbf{Pron.} & \textbf{Gram.} & \textbf{Task} & \textbf{Ethos} & \textbf{Tone} & \textbf{Str.} & \textbf{Weak.} & \textbf{Rec.} \\
\midrule
Debatrix & \cmark &  &  & \cmark & \pmark &  & \pmark & \pmark &  & \pmark &  &  & \pmark & \pmark & \pmark &  \\
InspireScore & \cmark &  &  & \cmark &  & \pmark &  & \cmark &  &  &  &  &  &  &  &  \\
Debatable Intelligence & \cmark &  &  & \pmark &  &  &  & \pmark &  & \pmark &  & \pmark & \pmark &  &  &  \\
MA Debate Chatbot (CT) & \cmark &  &  & \cmark & \pmark &  & \pmark & \pmark &  &  & \pmark &  &  & \pmark & \pmark &  \\
PolyDebate & \cmark & \cmark & \cmark & \cmark & \cmark & \cmark & \cmark & \cmark & \cmark & \cmark & \cmark & \cmark & \cmark & \cmark & \cmark & \cmark \\
\bottomrule
\end{tabular}}
\\[2pt]
\scriptsize
\cmark=covered; \pmark=partial; blank=not covered. T=text; A=audio; V=video. Arg.=argumentation/logos; Opp.=opponent targeting; Pathos=emotional/non-verbal appeal; Strat.=rhetorical/argumentative strategies; Org.=organization/relevance; Pron.=pronunciation/fluency; Gram.=grammar/lexis; Task=task fulfillment/stage appropriateness; Ethos=credibility/audience appeal; Tone=language/interpersonal appropriateness; Str.=strengths; Weak.=weaknesses; Rec.=recommendations.
\end{table}

\subsection{AI Judge Feedback Quality and Ablation}

We next evaluated whether the AI judge produces complete, specific, actionable, and grounded feedback. Using the same GPT-5.4 mini model, we compared the full PolyDebate judge against a generic baseline and four ablations that remove the rubric, the skill card, the multimodal evidence, or the structured feedback schema. The evaluation used 100 multimodal samples (transcript, audio, and video) whose weakness labels follow the ten subcategories in Table~\ref{tab:evaluation-coverage}.

Table~\ref{tab:judge-feedback-quality} shows the full judge performing best on all metrics: against the generic judge, it substantially improves weighted rubric coverage and yields more specific, actionable, and grounded feedback. The ablations confirm each component's contribution. Removing the rubric sharply reduces coverage, while removing the skill card or the multimodal evidence reduces weakness F1. Therefore, the full judge gives the most comprehensive formative diagnosis while preserving high feedback quality.

\begin{table}
\caption{AI judge feedback quality and component ablation. Coverage and weakness F1 are percentages; specificity, actionability, and groundedness use a 1--5 scale.}
\label{tab:judge-feedback-quality}
\centering
\resizebox{0.55\textwidth}{!}{%
\begin{tabular}{lccccc}
\toprule
\textbf{Variant} & \textbf{W.Cov.} & \textbf{W.F1} & \textbf{Spec.} & \textbf{Act.} & \textbf{Grd.} \\
\midrule
Generic LLM feedback judge & 46.7 & 80.6 & 4.0 & 4.1 & 4.1 \\
Full w/o rubrics & 51.3 & 64.3 & 4.4 & 4.5 & 4.7 \\
Full w/o skill card & 81.7 & 59.6 & 4.4 & 4.4 & 4.6 \\
Full w/o multimodal evidence & 71.7 & 32.2 & 4.2 & 4.2 & 4.1 \\
Full w/o feedback schema & 76.2 & 74.8 & 4.8 & 4.8 & 4.8 \\
Full PolyDebate judge & \textbf{99.2} & \textbf{85.9} & \textbf{4.9} & \textbf{4.9} & \textbf{4.9} \\
\bottomrule
\end{tabular}}
\\[2pt]
\scriptsize W.Cov.=weighted rubric coverage; W.F1=weakness-label F1; Spec.=specificity; Act.=actionability; Grd.=groundedness.
\end{table}

\subsection{User Perception Study}

Finally, we ran a lightweight study with ten students interested in English learning or debate. Each participant tried both the Unity 3D and web platform versions and then completed the same 5-point Likert questionnaire over six dimensions: usability, AI opponent usefulness, AI judge feedback usefulness, skill-card support, gameful motivation, and overall value.

\begin{figure}[t]
\centering
\includegraphics[width=0.65\linewidth]{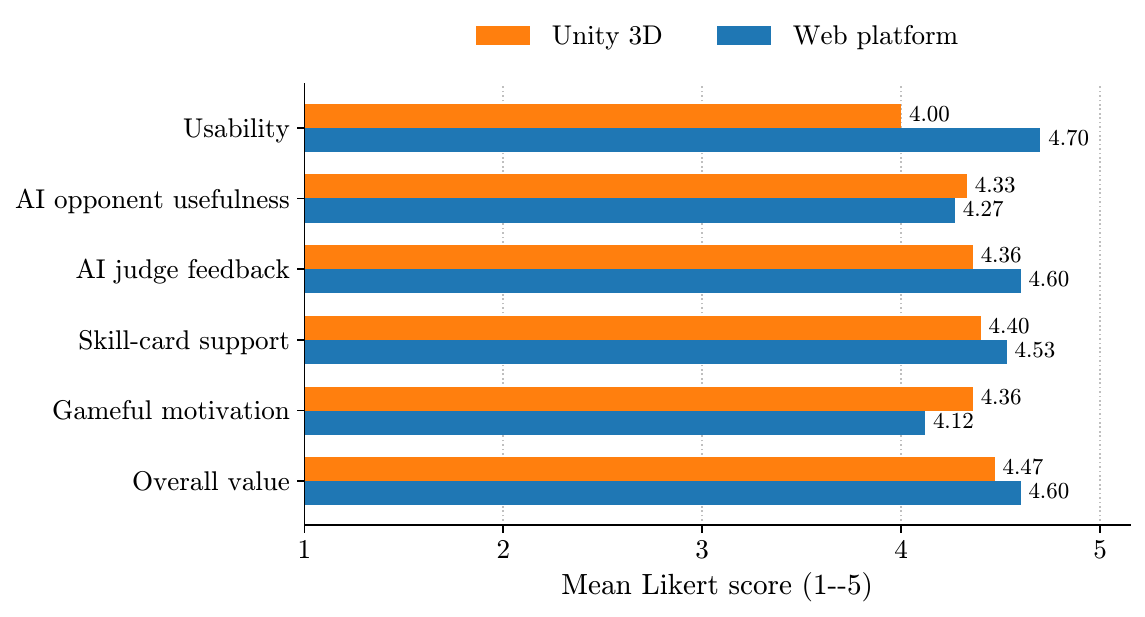}
\caption{User-perception results for the Unity 3D and web platform versions of PolyDebate. Bars report mean 5-point Likert ratings from ten students across six questionnaire dimensions.}
\label{fig:user-perception}
\end{figure}

Figure~\ref{fig:user-perception} summarizes the mean scores: both versions were rated positively on all dimensions. The web version scored higher on usability, feedback usefulness, skill-card support, and overall value, while the Unity version scored higher on gameful motivation, and AI opponent usefulness was similar across versions. The two are thus complementary: the web version suits accessible, repeated practice, while the Unity version supports immersive, game-like engagement.

\section{Conclusion and Future Work}

We presented PolyDebate, a game-orchestrated multimodal system for English debate practice and evaluation that connects staged 1v1 interaction, skill-card guidance, props and coins, an AI opponent, multimodal evidence, rubric-aligned evaluation, and structured feedback. PolyDebate is available in both an immersive Unity 3D version and a web platform version that share one workflow. Our evaluation shows that PolyDebate produces skill-aware opponent responses, broadens assessment beyond text-only evaluation, and presents the practice loop in a usable and motivating form.

Future work will explore learner-profile-driven practice that uses prior-round feedback, additional formats such as team-based and multi-role debate, and classroom deployment to examine learning gains.

\begin{credits}
\subsubsection{\ackname} This research work was conducted at the Research Institute for Artificial Intelligence of Things (RIAIoT) and The Institute for Higher Education Research and Development (IHERD) of PolyU. This work was supported in part by the Hong Kong Research Grants Council Theme-based Research Scheme (No. T43-518/24-N), PolyU Internal Research Fund (No. BDZ3), and PolyU LTC Project (Grant TDLEG25-28/IICA/P/05, No. 48EM).
\end{credits}

\bibliographystyle{splncs04}
\bibliography{references}

\end{document}